\RequirePackage{fix-cm}
\documentclass[smallcondensed]{svjour3}       
\smartqed  
\usepackage{graphicx}
\usepackage{enumerate}
\usepackage{amsmath}
 \usepackage{framed}
  \usepackage{amssymb}
  \usepackage[justification=centering]{caption}
\newtheorem{Theorem}{Theorem} 
\newtheorem{Definition}{Definition}
\newtheorem{Lemma}{Lemma}

\begin{document}
\title{Using Bernstein-Vazirani algorithm to attack block ciphers
}
\subtitle{}


\author{ Huiqin Xie$^{123}$      \and
         Li Yang$^{12}$
}

\authorrunning{Huiqin Xie, Li Yang} 

\institute{Huiqin Xie \\
              \email{xiehuiqin@iie.ac.cn}      \\     
           Li Yang \\
           \email{yangli@iie.ac.cn}\\
           \at
              1.State Key Laboratory of Information Security, Institute of Information Engineering, Chinese Academy of Sciences, Beijing, China \\
           \at
              2.Data Assurance and Communication Security Research Center,Chinese Academy of Sciences, Beijing, China\\
              \at
              3.School of Cyber Security, University of Chinese Academy of Sciences, Beijing, China }

\date{Received: date / Accepted: date}

\maketitle

\begin{abstract}
In this paper, we study applications of Bernstein-Vazirani algorithm and present several new methods to attack block ciphers. Specifically, we first present a quantum algorithm for finding the linear structures of a function. Based on it, we propose new quantum distinguishers for the 3-round Feistel scheme and a new quantum algorithm to recover partial key of the Even-Mansour construction. Afterwards, by observing that the linear structures of a encryption function are actually high probability differentials of it, we apply our algorithm to differential analysis and impossible differential cryptanalysis respectively. We also propose a new kind of differential cryptanalysis, called quantum small probability differential cryptanalysis, based on the fact that the linear structures found by our algorithm are also the linear structure of each component function. To our knowledge, no similar method was proposed before. The efficiency and success probability of all attacks are analyzed rigorously. Since our algorithm treats the encryption function as a whole, it avoid the disadvantage of traditional differential cryptanalysis that it is difficult to extending the differential path.
\end{abstract}

\section{Introduction}
\label{intro}
Over the last few years, there has been an increasing interest in quantum cryptography. On one hand, many cryptographic schemes based on quantum information have been proposed, among which the most well-known result is quantum key distribution (OKD) \cite{BB84}. These schemes take full advantage of the novel properties of quantum information and aims to realize functionalities that do not exist using classical information alone. On the other hand, the development of quantum computing threatens many classical cryptosystems. The most representative example is Shor's algorithm \cite{PS94}. By using Shor's algorithm, an adversary who owns a quantum computer can break the security of any schemes based on factorization or discrete logarithm, such as RSA. This has greatly motivated the development of post-quantum cryptography, i.e., classical cryptosystems that remain secure even when the adversary owns a quantum computer.

While currently used public-key cryptography suffers from a severe threat due to Shor's algorithm, the impact of quantum computers on symmetric-key cryptography is still less understood. Since Grover's algorithm provides a quadratic speed-up for general search problems, the key lengths of symmetric-key cryptosystems need to be doubled to maintain the security. In addition, Simon's algorithm \cite{SD97} has also been applied to cryptanalysis. Kuwakado and Morii use it to construct a quantum distinguisher for 3-round Feistel scheme \cite{KM10} and recover partial key of Even-Mansour construction \cite{KM12}. Santoli and Schaffiner extend their result and present a quantum forgery attack on CBC-MAC scheme \cite{SS17}. In \cite{KL16}, Kaplan et al. use Simon's algorithm to attack various symmetric cryptosystems, such as CBC-MAC, PMAC, CLOC and so on. They also study how differential and linear cryptanalysis behave in the post-quantum world \cite{KLL16}. In addition to Simon's algorithm, Bernstein-Vazirani (BV) algorithm \cite{BV93} has also been used for cryptanalysis. Li and Yang proposed two methods to execute quantum differential cryptanalysis based on BV algorithm in \cite{LY15}, but in their attack, it is implicitly assumed that the attacker can query the function which maps the plaintext to the input of the last round of the encryption algorithm.

In this paper, we study applications of BV algorithm and use it to attack block ciphers. It has been found that running BV algorithm on a Boolean function $f$ without performing the final measurement will gives a superposition of all states $|\omega\rangle$ ($\omega\in\{0,1\}^n$), and the amplitude corresponding to each $|\omega\rangle$ is its Walsh spectrum $S_f(\omega)$ \cite{FA13,HA11}. In addition, there is a link between the linear structure of a Boolean function and its Walsh spectrum \cite{DS01}. Based on these two facts, Li and Yang present a quantum algorithm to find the linear structures of a Boolean function in \cite{LY16}. We modify their algorithm so that it can find the linear structures of a vector function. Our attack strategies are all built on this modified algorithm.

\textbf{Attack model.} In this paper, we only consider quantum chosen message attack that has been studied in \cite{BZ13,DF13,TAC16}. In this attack model, the adversary is granted the access to a quantum oracle which computes the encryption function in superposition. Specifically, if the encryption algorithm is described by a classical function $E_k:\{0,1\}^n\rightarrow\{0,1\}^n$, then the adversary can make quantum queries $\sum_{x,y}|x\rangle|y\rangle\rightarrow\sum_{x,y}|x\rangle|y\oplus E_k(x)\rangle $.

\textbf{Our contributions.} In this article, we present several methods to attack block ciphers. We first propose a quantum algorithm for finding the linear structures of a vector function, which takes BV algorithm as a subroutine and is developed from the algorithm in \cite{LY16}. Then we modify this original algorithm to get different versions and apply them in different attack strategies. In more detail, our main contributions are as follows:
\vskip 0.2cm

\noindent
$\bullet\,$ We construct new quantum distinguishers for the 3-round Feistel scheme and propose a new quantum algorithm to recover partial key of Even-Mansour construction. Our methods are similar with the ones proposed by Kuwakado and Morii \cite{KM10,KM12}, but we use BV algorithm instead of Simon's algorithm. Although this modification cause a slight increase in complexity, it makes our methods has more general applications. For example, by constructing functions that have different linear structures, we can obtain various distinguishers for the 3-round Feistel scheme. The essential reason is that using BV algorithm can find not only the periods of a function but also its other linear structures.
\vskip 0.2cm

\noindent
$\bullet\,$ Observing that linear structures of a encryption function are actually high probability differentials of it, we propose three ways to execute differential cryptanalysis, which we call quantum differential analysis, quantum small probability differential cryptanalysis and quantum impossible differential cryptanalysis respectively. Afterwards, we analyze the efficiency and success probability of all attacks. The quantum algorithms used for these three kinds of differential cryptanalysis all have polynomial running time. As we know, one of the main shortcomings of traditional differential cryptanalysis is the difficulties in extending the differential paths, which limits the number of rounds that can be attacked. Our approach avoids this problem since it treats the encryption function as a whole.

\section{Preliminaries}
In this section, we briefly recall a few notations and results about the linear structure. Let $n$ be a positive integer and $\mathbb{F}_2=\{0,1\}$ be a finite field of characteristic 2. $\mathbb{F}_2^n=\{0,1\}^n$ is a vector space over $\mathbb{F}_2$. The set of all functions from $\mathbb{F}_2^n$ to $\mathbb{F}_2$ is denoted as $\mathcal{B}_n$.
\begin{Definition}
A vector $a\in \mathbb{F}_2^n$ is called a linear structure of a Boolean function $f\in\mathcal{B}_n$, if
$$
f(x\oplus a)+f(x)=f(a)+f(0),\,\,\,\forall x\in \mathbb{F}_2^n,
$$
where $\oplus$ denotes the bitwise exclusive-or.
\end{Definition}

For any $f\in\mathcal{B}_n$, let $U_f$ be the set of all linear structures of $f$, and
$$
U_f^i=\{a\in \mathbb{F}_2^n|f(x\oplus a)+f(x)=i,\forall x\in \mathbb{F}2^n\},\,\,\, i=0,1.
$$
Obviously, $U_f=U_f^0\cup U_f^1$. For any $a\in \mathbb{F}_2^n$ and $i=0,1$, let
$$
V_{f,a}^i=\{x\in \mathbb{F}_2^n|f(x\oplus a)+f(x)=i\}.
$$
Clearly, $0\leq|V_{f,a}^i|/2^n\leq 1$, and $a\in U^i_f$ if and only if $|V_{f,a}^i|/2^n=1$. For any $a\in \mathbb{F}_2^n$, $1-|V_{f,a}^i|/2^n$ quantifies how close a is to be a linear structure of $f$. Naturally, we give the following definition:
\begin{Definition}A vector $a\in \mathbb{F}_2^n$ is called a $\sigma$-close linear structure of a function $f\in\mathcal{B}_n$, if there exists $i\in\{0,1\}$ such that
$$
1-\frac{|\{x\in \mathbb{F}_2^n|f(x\oplus a)+f(x)=i\}|}{2^n}<\sigma.
$$
\end{Definition}

If $a$ is a $\sigma(n)$-close linear structure of $f$ for some negligible function $\sigma(n)$, we call it a quasi linear structure of $f$.

Relative differential uniformity of a Boolean function quantifies how close the function is from having a nontrivial linear structure, which is defined in \cite{NK93}:
\begin{Definition}The relative differential uniformity of $f\in\mathcal{B}_n$ is defined as
$$
\delta_f=\frac{1}{2^n}\max_{0\neq a\in \mathbb{F}_2^n}\max_{i\in \mathbb{F}_2}|\{x\in \mathbb{F}_2^n|f(x\oplus a)+f(x)=i\}|.
$$
\end{Definition}
For any $f\in\mathcal{B}_n$, it is obviously that $\frac{1}{2}\leq\delta_f\leq1$, and $U_f\neq\{0\}$ if and only if $\delta_f=1$. The linear structures of a Boolean function is closely related to its Walsh spectrum, which is defined as follows:
\begin{Definition} The Walsh spectrum of a Boolean function $f\in\mathcal{B}_n$ is defined as
$$
S_f(\omega)=\frac{1}{2^n}\sum_{x\in \mathbb{F}_2^n}(-1)^{f(x)+\omega\cdot x}.
$$
\end{Definition}

The relation between the linear structure and Walsh spectral is captured by following two lemmas, which have been proved in \cite{NK90}.
\begin{Lemma}Let $f\in\mathcal{B}_n$, then $\forall a\in \mathbb{F}_2^n$, $\forall i\in \mathbb{F}_2$,
$$
\sum_{\omega\cdot a=i}S^2_f(\omega)=\frac{|V^i_{f,a}|}{2^n}=\frac{|\{x\in \mathbb{F}_2^n|f(x\oplus a)+f(x)=i\}|}{2^n}.
$$
\end{Lemma}
\begin{Lemma}For $f\in\mathcal{B}_n$, let $N_f=\{\omega\in \mathbb{F}_2^n|S_f(\omega)\neq0\}$. Then $\forall i\in\{0,1\}$, it holds that
$$
U_f^i=\{a\in \mathbb{F}_2^n|\omega\cdot a=i,\forall\,\, \omega\in N_f\}.
$$
\end{Lemma}

Lemma 2 provides a method to find the linear structures. If we have a sufficiently large subset $H$ of $N_f$, we can get $U_f^i$ by solving the system of linear equations $\{a\cdot\omega=i|\omega\in H\}$. (Here, solving the system of linear equations $\{a\cdot\omega=i|\omega\in H\}$ means finding vector $a$ that satisfies $a\cdot \omega=i$ for all $\omega\in H$.)

Next we consider the vector functions. Suppose $m,n$ are positive integers. $\mathcal{C}_{m,n}$ denotes the set of all functions from $\mathbb{F}_2^m$ to $\mathbb{F}_2^n$. The linear structure of a vector function in $\mathcal{C}_{m,n}$ can be defined similarly:
\begin{Definition}A vector $a\in \mathbb{F}_2^m$ is called a linear structure of a vector function $F\in\mathcal{C}_{m,n}$, if there exists a vector $\alpha\in\{0,1\}^n$ such that
$$
F(x\oplus a)\oplus F(x)=\alpha, \,\,\,\forall x\in\{0,1\}^m.
$$
\end{Definition}

Suppose $F=(F_1,F_2,\cdots,F_n)$. A straightforward way to find the linear structures of $F$ is to first search for the linear structures of each component function $F_j$ respectively and then take the intersection. Let $U_F$ be the set of all linear structures of $F$, and $U_F^{\alpha}=\{a\in \mathbb{F}_2^n|F(x\oplus a)\oplus F(x)=\alpha, \,\,\forall x\}$. It is obviously that $U_F=\cup_{\alpha}U_F^{\alpha}$. The relative differential uniformity of $F$ is defined as
$$
\delta_F=\frac{1}{2^m}\max_{0\neq a\in \mathbb{F}_2^m}\max_{\alpha\in \mathbb{F}_2^n}|\{x\in \mathbb{F}_2^m|F(x\oplus a)\oplus F(x)=\alpha\}|,
$$
which quantifies how close $F$ is from having a nontrivial linear structure.

\section{Finding linear structures via Bernstein-Vazirani algorithm  }
In this section we briefly recall the BV algorithm \cite{BV93} and introduce how to use it to find the linear structures of a Boolean function. The goal of BV algorithm is to determine a secret string $a\in\{0,1\}^n$. Specifically, suppose
$$
f(x)=a\cdot x=\sum_{i=1}^na_ix_i.
$$
The algorithm aims to determine $a$, given the access to an quantum oracle which computes the function $f$. It works as follows:
\begin{enumerate}[  1.]
\item Prepare the initial state $|\psi_0\rangle=|0\rangle^{\otimes n}|1\rangle$, then perform the Hadamard transform $H^{(n+1)}$ on it to obtain the quantum superposition
$$
|\psi_1\rangle=\sum_{x\in \mathbb{F}_2^n}\frac{|x\rangle}{\sqrt{2^n}}\cdot\frac{|0\rangle-|1\rangle}{\sqrt{2}}.
$$

\item A quantum query to the oracle which computes $f$ maps it to the state
$$
|\psi_2\rangle=\sum_{x\in\mathbb{F}_2^n}\frac{(-1)^{f(x)}|x\rangle}{\sqrt{2^n}}\cdot\frac{|0\rangle-|1\rangle}{\sqrt{2}}.
$$

\item Apply the Hadamard gates $H^{(n)}$ to the first $n$ qubits again yielding
$$
|\psi_3\rangle=\sum_{y\in \mathbb{F}_2^n}\frac{1}{2^n}\sum_{x\in \mathbb{F}_2^n}(-1)^{f(x)+y\cdot x}|y\rangle,
$$
where we omit the last qubit for the simplicity. If $f(x)=a\cdot x$, we have
\begin{align*}
|\psi_3\rangle&=\sum_{y\in \mathbb{F}_2^n}(\frac{1}{2^n}\sum_{x\in \mathbb{F}_2^n}(-1)^{(a\oplus y)\cdot x})|y\rangle\\
&=\sum_{y\in \mathbb{F}_2^n}\delta_{a}(y)|y\rangle\\
&=|a\rangle,
\end{align*}
where $\delta_a(y)=1$ if $y=a$, otherwise $\delta_a(y)=0$. Then by measuring $|\psi_3\rangle$ in the computational basis, we will get $a$ with probability 1.
\end{enumerate}

If we run the BV algorithm on a general function $f\in\mathcal{B}_n$, the output before the measurement can be expressed as
$$
|\psi_3\rangle=\sum_{y\in \mathbb{F}_2^n}S_f(y)|y\rangle,
$$
where $S_f(\cdot)$ is the Walsh spectrum of $f$. When we measure the above state in the computational basis, we will obtain $y$ with probability $S_f(y)^2$. In other words, we will always get $y\in N_f$ when we run the BV algorithm on $f$. This fact combined with Lemma 2 implies a way to find the linear structures.

Now we state the quantum algorithm proposed in \cite{LY16} for finding the linear structures of a Boolean function. Roughly speaking, the BV algorithm is treated as a subroutine. By repeating the subroutine until one gets a subset $H$ of $N_f$, and then solving the system of linear equations $\{x\cdot\omega=i|\omega\in H\}$ for both $i=0$ and $1$, one will get candidate linear structures of $f$.

\begin{framed}
\noindent
\textbf{Algorithm 1}

\noindent
Let $p(n)$ be an arbitrary polynomial function of $n$. $\Phi$ denotes the null set. Initialize the set $H:=\Phi$.\\
1 \textbf{For} $p=1,2,\cdots,p(n)$, \textbf{do}

\noindent
\hangafter 1
\hangindent 2.7em
2 \quad $\,\,\,$ Run the BV algorithm with queries on the quantum oracle of $f$ to get an $n$-bit output $\omega\in N_f$.

\noindent
3 \quad $\,\,\,$ Let $H=H\cup \{\omega\}$.\\
4 \textbf{end}

\noindent
\hangafter 1
\hangindent 0.9em
5 Solve the system of linear equations $\{x\cdot \omega=i|\omega\in H\}$ to get solution $A^i$ for $i=0,1$ respectively.

\noindent
6 \textbf{If} $A^0\cup A^1\subseteq\{(0,\cdots,0)\}$, \textbf{then} output ``No'' and halt.\\
7 \textbf{Else}, output $A^0$ and $A^1$.
\end{framed}

To justify the validity of the above algorithm, we present following two theorems, where the Theorem 1 is proved in \cite{LY16} and the Theorem 2 hasn't been proved before.

\begin{Theorem}If running Algorithm 1 on a function $f\in\mathcal{B}_n$ gives sets $A^0$ and $A^1$, then for all $a\in A^i$ ($i=0,1$), all $\epsilon$ satisfying $0<\epsilon<1$, we have
\begin{equation}
Pr\Big[1-\frac{|\{x\in \mathbb{F}_2^n|f(x\oplus a)+f(x)=i\}|}{2^n}<\epsilon\Big]>1-\exp(-2p(n)\epsilon^2).
\end{equation}
\end{Theorem}

This theorem is proved in \cite{LY16}, and we present the proof in Appendix A for the paper to be self-contained.

For arbitrary function $f\in\mathcal{B}_n$, we let
$$
\delta'_f=\frac{1}{2^n} \max_{\substack{a\in \mathbb{F}_2^n\\ a\notin U_f}}\max_{i\in \mathbb{F}_2}|\{x\in \mathbb{F}_2^n|f(x\oplus a)+f(x)=i\}|.
$$
It is obviously that $\delta'_f<1$. And if $\delta_f<1$, it holds that $\delta'_f=\delta_f$. By the definition of $\delta'_f$, we can see that the smaller $\delta'_f$ is, the better for ruling out the vectors which are not linear structures of $f$ during executing Algorithm 1.
\begin{Theorem}Suppose $\delta'_f\leq p_0<1$ and Algorithm 1 makes quantum queries for $p(n)=cn$ times. Then it holds that\\
\vskip -0.03cm

\noindent
1.If $\delta_f<1$, that is, $f$ has no nonzero linear structure, then Algorithm 1 returns ``No'' with probability greater than $1-p_0^{cn}$;\\
\vskip -0.03cm

\noindent
2.If Algorithm 1 returns $A^0$ and $A^1$, then for any $a\notin U_f^i$ ($i=0,1$),
$$
Pr[a\in A^i]\leq p_0^{cn}.
$$
\end{Theorem}

\noindent
\textbf{\emph{Proof}}. We first prove the second conclusion. Without loss of generality, we suppose $i=0$. The case where $i=1$ can be proved by similar way. If $a\notin U_f^0$, then according to Lemma 2 there exists a vector $\omega\in N_f$ such that $\omega\cdot a=1$. Let $K=\{\omega\in N_f|\omega\cdot a=1\}$. If the $cn$ times of running BV algorithm ever gives a vector $\omega\in K$, then $a\notin A^0$. Let $W$ denote the random variable obtained by running BV algorithm, then
\begin{align*}
Pr[W\in K]&=\sum_{a\cdot\omega=1}S_f(\omega)^2\\
&=1-\frac{|V_{f,a}^0|}{2^n}\\
&\geq1-p_0.
\end{align*}
The second formula holds due to Lemma 1. Therefore,
\begin{align*}
P[a\in A^0]&=[1-P(W\in K)]^{cn}\\
&\leq p_0^{cn},
\end{align*}
which completes the proof of the second conclusion. By observing that $\delta_f<1$ means there is no nonzero vector in $U_f$, the first conclusion can be naturally derived from the second one.

$\hfill{} \Box$

Suppose $l(n)$ is an arbitrary polynomial of $n$. Theorem 1 implies that after $p(n)=O(l(n)^2n)$ queries, all vectors in $A^0$ and $A^1$ will be $\frac{1}{l(n)}$-close linear structures of $f$ except a negligible probability. In other words, Algorithm 1 is very likely to output the high probability differentials of $f$. Theorem 2 shows that if $f$ has no linear structure, Algorithm 1 with $O(n)$ queries will output ``No'' except a negligible probability. In addition, if Algorithm 1 returns sets $A^0$ and $A^1$, then each vector in $A^i$ will be linear structure of $f$ with overwhelming probability. (The probability of a event happening is said to be overwhelming if it happens except a negligible probability.)

\section{Linear structure attack}
In this section, we first improve the Algorithm 1 so that it can find the linear structures of a vector function. Afterwards, we use the new algorithm to construct quantum distinguishers for the 3-round Feistel scheme and recover partial key of Even-Mansour construction respectively. Since our attack strategy is based on the linear structures of some constructed functions, we call it linear structure attack.

\subsection{Attack algorithm}
Suppose $F=(F_1,F_2,\cdots,F_n)\in\mathcal{C}_{m,n}$. A straightforward way to find the linear structures of $F$ is to apply Algorithm 1 to each component function $F_j$ respectively and then choose a public linear structure. Specifically, we have following algorithm:
\begin{framed}
\noindent
\textbf{Algorithm 2}

\noindent
The access to the quantum oracle of $F=(F_1,\cdots,F_n)$ is given. $p(n)$ is an arbitrary polynomial function of $n$.\\
1 \textbf{For} $j=1,2,\cdots,n$, \textbf{do}

\noindent
\hangafter 1
\hangindent 2em
2 \quad Run Algorithm 1 with $p(n)$ queries on $F_j$ to get $A_j^0$ and $A_j^1$.

\noindent
3 \quad $\,$\textbf{If} Algorithm 1 outputs ``No'', \textbf{then} output ``No'' and halt.\\
4 \quad $\,$\textbf{Else} let $A_j= A_j^0\cup A_j^1$.\\
4 $\,$\textbf{end}\\
5  $\,$\textbf{If} $A_1\cap\cdots\cap A_n\subseteq\{(0,\cdots,0)\}$, \textbf{Then} output ``No'' and halt.

\noindent
\hangafter 1
\hangindent 1.11em
6 \textbf{Else} choose an arbitrary nonzero vector $a\in A_1\cap\cdots\cap A_n$ and output $(a,i_1,\cdots,i_n)$, where $i_1,\cdots,i_n$ is the superscript such that $a\in A_1^{i_1}\cap A_2^{i_2}$ $\cap\cdots\cap A_n^{i_n}$.
\end{framed}

For any function $F=(F_1,\cdots,F_n)$, let $\delta'_F=\max_{1\leq j\leq n}\delta'_{F_j}$. The following theorem justifies the validity of the Algorithm 2.
\begin{Theorem}Suppose $F\in\mathcal{C}_{m,n}$. Running Algorithm 2 with $cn^2$ queries ($p(n)=cn$) on $F$ gives ``No'' or some vector. It holds that\\
\vskip -0.03cm

\noindent
1.If $\delta'_F\leq p_0<1$ and $F$ has no linear structure, then Algorithm 2 returns ``No'' with probability greater than $1-p_0^{cn}$.\\
\vskip -0.03cm

\noindent
2.If $\delta'_{F}\leq p_0<1$, then for any $a\notin U_F^{(i_1,\cdots,i_n)}$, we have
$$
Pr[\text{Algorithm 2 returns } (a,i_1,\cdots,i_n)]\leq p_0^{cn}.
$$
\vskip -0.03cm

\noindent
3.If  $(a,i_1,\cdots,i_n)$ is obtained by running Algorithm 2, then for any $0<\epsilon<1$,
\begin{small}
\begin{equation}
Pr[\frac{|\{x\in \mathbb{F}_2^m|F(x\oplus a)\oplus F(x)=i_1\cdots i_n\}|}{2^m}>1-n\epsilon]>\big(1-\exp(-2p(n){\epsilon}^2)\big)^n.
\end{equation}
\end{small}
\end{Theorem}

\quad\\
\vskip -0.17cm

\noindent
\textbf{\emph{Proof}}. By observing the fact that $a\in U_F^{(i_1,\cdots,i_n)}$ if and only if for all $j=1,\cdots,n$, $a\in U_{F_j}^{i_j}$, the first two conclusion can be naturally derived from the Theorem 2. According to Theorem 1, we have
\begin{equation}
\frac{|\{x\in \mathbb{F}_2^m|F_j(x\oplus a)\oplus F_j(x)=i_j\}|}{2^m}>1-\epsilon,\,\,\, \forall j=1,\cdots,n
\end{equation}
holds with a probability greater than $(1-\exp(-2p(n){\epsilon}^2))^n$. If the equation $(3)$ holds, then the number of $x$ satisfying
\begin{equation}
F_j(x\oplus a)+F_j(x)=i_j
\end{equation}
for both $j=1$ and $j=2$ is at least $2^m[2(1-\epsilon)-1]=2^m(1-2\epsilon)$. Similarly, the number of $x$ satisfying the equation $(4)$ for all $j=1,2,3$ is at least $2^m[(1-2\epsilon)+(1-\epsilon)-1]=2^m(1-3\epsilon)$. By induction, the number of $x$ satisfying $(4)$ for all $j=1,\cdots,n$ is at least $2^m(1-n\epsilon)$. Thus the probability that
$$
\frac{|\{x\in \mathbb{F}_2^m|F(x\oplus a)\oplus F(x)=i_1\cdots i_n\}|}{2^m}>1-n\epsilon
$$
holds is greater than $(1-\exp(-2p(n){\epsilon}^2))^n$. Thus the third conclusion holds.

$\hfill{} \Box$

Note that Algorithm 2 actually requires that the adversary has the oracle access to each component function of $F$. About the efficiency, since Algorithm 2 needs to find the intersection of the sets $A_j\,'s$, its complexity depends on the size of these sets, which relies on the properties of the specific function $F$. However, we can prove that only polynomial time of computation is needed when Algorithm 2 is applied to 3-round Feistel scheme or Even-Mansour construction. In \cite{KL16,KM10,KM12,SS17}, the authors use Simon's algorithm to find the period of some constructed functions and then break the security of 3-round Feistel scheme or Even-Mansour construction. Compared with Simon's algorithm, the complexity of Algorithm 2 is a little larger because it needs to search linear structures of each component function respectively. However, Algorithm 2 has more general applications. It can find not only the periods of a function but also its other linear structures, which allows us to construct multiple distinguishers for 3-round Feistel scheme. And in the section 5 we will see that such way of finding linear structures by considering each component function respectively may bring some unexpected advantages for differential cryptanalysis.


\subsection{Application to a three-round Feistel scheme}
A Feistel scheme is a classical construction to build block ciphers. A 3-round Feistel scheme with input $(x_L,x_R)$ and output $(y_L,y_R)$ is built from three random functions $P_1,P_2,P_3$ as shown in Figure 1, where $x_L,x_R,y_L,y_R\in\{0,1\}^n$. It's proved that a 3-round Feistel scheme is a secure pseudorandom permutation as long as the internal functions are pseudorandom as well \cite{LR88}. Our goal is to construct a quantum distinguisher which distinguishes a 3-round Feistel scheme from a random permutation on $\{0,1\}^{2n}$.
\begin{figure}
  \centering
  \includegraphics[width=3cm]{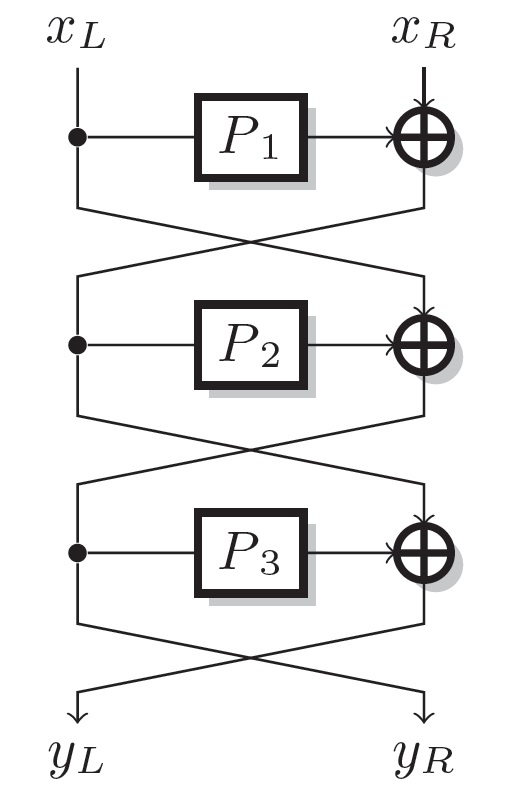}
  \caption{Three-round Feistel scheme}\label{1}
\end{figure}

Suppose $s_0,s_1\in \mathbb{F}_2^n$ are two arbitrary constants such that $s_0\neq s_1$. We define the following function:
\begin{align}
F: \mathbb{F}_2\times \mathbb{F}_2^n&\rightarrow \mathbb{F}_2^n\nonumber\\
(\,\,\,b\,\,,\,x\,\,)&\rightarrow P_2(x\oplus P_1(s_b)).
\end{align}
Given the oracle access of the 3-round Feistel function $E$, it is easy to construct the oracle $O_F$ which computes $F$ on superpositions. Observing that the right part of the output $E(s_b,x)$ is $F(b,x)\oplus s_b$, we can construct the oracle $O_F$ by first querying the oracle which computes the right part of $E$, then applying the unitary operator: $U^{s_0,s_1}:|b\rangle|c\rangle\rightarrow|b\rangle|c\oplus s_b\rangle$. For every $(b\|x)\in \mathbb{F}_2^{n+1}$, it is easy to check that
$$
F(b,x)=F(b\oplus1,x\oplus P_1(s_0)\oplus P_1(s_1)).
$$
Thus $(1\|s)\triangleq(1\|P_1(s_0)\oplus P_1(s_1))$ is a nonzero linear structure of $F$, or more accurately, $(1\|s)\in U_F^{(0,\cdots,0)}$. Therefore, by running Algorithm 2 on $F$ one can get $(1\|s)$. On the other hand, the probability of a random function having a linear structure is negligible. Given the access to a quantum oracle which computes the 3-rounds Feistel function $E$ or a random permutation over $\{0,1\}^{2n}$, we can construct the distinguishing algorithm as below:
\begin{framed}
\noindent
\textbf{Algorithm 3}

\noindent
Let $p(n)=n+1$ and initialize the set $H:=\Phi$.\\
1 Choose $s_0,s_1\in \mathbb{F}_2^n$, s.t. $s_0\neq s_1$. Then define function $F=(F_1,\cdots,F_n)$ as equation $(5)$.\\
2 \textbf{For} $j=1,\cdots,n$ \textbf{do}\\
3 \quad\textbf{For} $p=1,\cdots,p(n)$, \textbf{do}\\
4 \quad\quad Run BV algorithm on $F_j$ to get an output $\omega\in N_{F_j}$.\\
5 \quad\quad Let $H=H\cup\{\omega\}$\\
6 \quad\textbf{End}

\noindent
\hangafter 1
\hangindent 1.7em
7 \quad Solve the system of linear equations $\{x\cdot \omega=0|\omega\in H\}$ to get the solution $A_j^0$.

\noindent
8 \quad \textbf{If} $A_j^0\subseteq\{(0,\cdots,0)\}$, \textbf{then} output ``No'' and halt.\\
9 \quad \textbf{Else} Let $H=\Phi$.\\
10 \textbf{End}\\
11 \textbf{If} $A_1^0\cap\cdots\cap A_{n+1}^0\subseteq\{(0,\cdots,0)\}$, \textbf{then} output ``No'' and halt.

\noindent
\hangafter 1
\hangindent 1.5em
12 \textbf{Else} choose an arbitrary nonzero vector $a\in A_1^0\cap\cdots\cap A_{n+1}^0$. And choose an $(n+1)$-bit string $u$ uniformly at random. Let $u'=u\oplus a$. Make classical queries for $u$ and $u'$.

\noindent
13 \textbf{If} $F(u)=F(u')$, \textbf{Then} output ``Yes''.\\
14 \textbf{Else}, output ``No''.
\end{framed}

Before analyzing the validity and efficiency of Algorithm 3, we first give following lemma:

\begin{Lemma}
Suppose $F=(F_1,\cdots,F_{n})$ is defined as equation $(5)$. Then for all $j=1,2,\cdots,n$,
\begin{small}
$$
\delta_{F_j}(1\|s)\,\,\,\triangleq\frac{1}{2^{n+1}}\max_{\substack{(\tau,t)\in \mathbb{F}_2^{n+1}\\ (\tau,t)\notin\{(0,\cdots,0),(1\|s)\}}}\,\,\,|\{(b,x)\in \mathbb{F}_2^{n+1}|F_j(b,x)=F_j(b\oplus\tau,x\oplus t)\}|\leq\frac{2}{3}
$$
\end{small}
holds except a negligible probability. Here, $\delta_{F_j}(1\|s)$ is still a random variable since $F_j$ is determined by random functions $P_1, P_2$.
\end{Lemma}

\noindent
\textbf{\emph{Proof}}. If $\delta_{F_j}(1\|s)>\frac{2}{3}$, then there exists $(\tau,t)\notin\{0,(1\|s)\}$ such that $|\{(b,x)\in \mathbb{F}_2^{n+1}|F_j(b,x)=F_j(b\oplus \tau,x\oplus t)\}|>\frac{2}{3}\cdot 2^{n+1}$. If $\tau=0$, this implies
\begin{small}
$$
|\{x\in \mathbb{F}_2^n|F_j(0,x)=F_j(0,x\oplus t)\}|>\frac{2}{3}\cdot 2^n\text{ or }|\{x\in \mathbb{F}_2^n|F_j(1,x)=F_j(1,x\oplus t)\}|>\frac{2}{3}\cdot2^n.
$$
\end{small}
Thus there exists some $b$ such that
$$
|\{x\in \mathbb{F}_2^n|P_{2j}(x\oplus P_1(s_b))=P_{2j}(x\oplus t\oplus P_1(s_b))\}|>\frac{2}{3}\cdot2^n,
$$
where $P_{2j}$ is the $j^{th}$ component function of $P_2$. That is,
$$
|\{x\in \mathbb{F}_2^n|P_{2j}(x)=P_{2j}(x\oplus t)\}|>\frac{2}{3}\cdot2^n.
$$
Similarly, if $\tau=1$, we have
$$
|\{x\in \mathbb{F}_2^n|P_{2j}(x)=P_{2j}(x\oplus t\oplus P_1(s_0)\oplus P_1(s_1))\}|>\frac{2}{3}\cdot2^n.
$$
Anyway, there exists a $u\neq (0,\cdots,0)$ such that
\begin{equation}
|\{x\in \mathbb{F}_2^n|P_{2j}(x)=P_{2j}(x\oplus u)\}|>\frac{2}{3}\cdot2^n.
\end{equation}
Since $P_{2}$ is a random function, $P_{2j}$ is actually a random Boolean function. Thus, $|\{x\in \mathbb{F}_2^n|P_{2j}(x)=P_{2j}(x\oplus u)\}|$ is still a random variable. We next prove that equation $(6)$ holds only with a negligible probability. This will imply that the probability of $\delta_{F_j}(1\|s)>\frac{2}{3}$ is negligible, and thus completes the proof.

The rest is to prove the probability
$$
Pr[\frac{|\{x\in \mathbb{F}_2^n|P_{2j}(x)=P_{2j}(x\oplus u)\}|}{2^n}>\frac{2}{3}]
$$
is negligible, where $P_{2j}$ is uniformly chosen from the set of all Boolean functions from $\mathbb{F}_2^n$ to $\mathbb{F}_2$. We call an unordered set of two vectors in $\mathbb{F}_2^n$ a \emph{pair}. For example, vectors $x,y$ form a pair $\{x,y\}$, and $\{x,y\}=\{y,x\}$. The difference of a pair $\{x,y\}$ is defined as $x\oplus y$. For the nonzero difference $u$, there are $2^{n-1}$ pairs with this difference. Thus,
\begin{align}
&\big|\big\{x\in \mathbb{F}_2^n|P_{2j}(x)=P_{2j}(x\oplus u)\big\}\big|\notag\\
=&\big|\big\{(x,y)\in \mathbb{F}_2^n\times \mathbb{F}_2^n|P_{2j}(x)=P_{2j}(y),x\oplus y=u\big\}\big|\notag\\
=&2\big|\big\{\{x,y\}|P_{2j}(x)=P_{2j}(y),x\oplus y=u\big\}\big|.
\end{align}
For convenience, we denote the $2^{n-1}$ pairs with difference $u$ as $\{x_1,x_1\oplus u\},\{x_2,x_2\oplus u\},\cdots,\{x_{2^{n-1}},x_{2^{n-1}}\oplus u\}$. Let $Z_l=P_{2j}(x_l)\oplus P_{2j}(x_l\oplus u)\oplus1$ for $l=1,2,\cdots, 2^{n-1}$. Since $P_{2j}$ is random Boolean function and for $l\neq k$, $\{x_l,x_l\oplus u\}\bigcap\{x_k,x_k\oplus u\}=\Phi$, $Z_1,\cdots, Z_{2^{n-1}}$ are independent and identically distributed random variables. They all follow the uniform distribution over $\{0,1\}$. That is, $Pr[Z_l=0]=Pr[Z_l=1]=\frac{1}{2}$, $l=1,2,\cdots,2^{n-1}$. According to Hoeffding's inequality,
$$
Pr[\frac{1}{2^{n-1}}\sum_{l=1}^{2^{n-1}}Z_l-\frac{1}{2}\geq\frac{1}{6}]\leq2\exp(-\frac{1}{18}2^{n-1}).
$$
That is,
$$
Pr[\sum_{l=1}^{2^{n-1}}Z_l\geq\frac{2}{3}\cdot 2^{n-1}]\leq2\exp(-\frac{1}{18}2^{n-1}).
$$
Since
\begin{align*}
\sum_{l=1}^{2^{n-1}}Z_l=&\big|\big\{Z_l|Z_l=1\big\}\big|\\
=&\big|\big\{\,\{x_l,x_l\oplus u\}\,|P_{2j}(x_l)\oplus P_{2j}(x_l\oplus u)\oplus1=1\big\}\big|\\
=&\big|\big\{\,\{x,y\}\,|P_{2j}(x)=P_{2j}(y),x\oplus y=u\big\}\big|,
\end{align*}
we have
$$
Pr\Big[\,\big|\big\{\,\{x,y\}\,|P_{2j}(x)=P_{2j}(xy),x\oplus y=u\big\}\big|\geq\frac{2}{3}\cdot2^{n-1}\Big]\leq2\exp(-\frac{1}{18}2^{n-1}).
$$
According to equation $(7)$, we have
$$
Pr\left[\frac{|\{x\in \mathbb{F}_2^n|P_{2j}(x)=P_{2j}(x\oplus u)\}|}{2^n}>\frac{2}{3}\right]\leq2\exp(-\frac{1}{18}2^{n-1}).
$$
Therefore, equation $(6)$ holds with a negligible probability.

$\hfill{} \Box$

About the validity of Algorithm 3, we have following theorem:
\begin{Theorem} Algorithm 3 successfully distinguishes the 3-round Feistel function from a random permutation except a negligible probability.
\end{Theorem}

\noindent
\textbf{\emph{Proof}}. If the given oracle computes a random permutation, the string $a$ obtained during executing Algorithm 3 is random if exists. Hence the probability of $F(u)$ being equal to $F(u')$ is approximate to $\frac{1}{2^n}$. On the other hand, if the given oracle computes the 3-round Feistel function, then
\begin{small}
$$
\delta_{F_j}(1\|s)\,\,\,\triangleq\frac{1}{2^{n+1}}\max_{\substack{(\tau,t)\in \mathbb{F}_2^{n+1}\\ (\tau,t)\notin\{(0,\cdots,0),(1\|s)\}}}\,\,\,|\{(b,x)\in \mathbb{F}_2^{n+1}|F_j(b,x)=F_j(b\oplus\tau,x\oplus t)\}|\leq\frac{2}{3}
$$
\end{small}

\noindent
holds with a overwhelming probability according to Lemma 3. Due to Theorem 2, above equation indicates
$$
Pr[a\neq (1\|s)]\leq(\frac{2}{3})^{n+1}.
$$
Thus the probability that $F(u)\neq F(u')$ is no more than $(\frac{2}{3})^{n+1}$, which completes the proof.

$\hfill{} \Box$

\noindent

Note that Algorithm 3 actually requires that the attacker can query each component function of the right part of $E$. Then we consider the efficiency of Algorithm 3. If the given oracle computes the 3-round Feistel function, according to Lemma 3 and Theorem 2, for any $a\notin\{(0,\cdots,0),(1\|s)\}$, we have $Pr[a\in A_j^0]\leq(\frac{2}{3})^{n+1}$. Thus with a overwhelming probability, $A_j^0$ contains only $0$ and $(1\|s)$. Therefore, finding the intersection of $A_j^0\,'s$ almost needs no calculation. It is also true when given oracle computes a random permutation. In addition, Algorithm 3 queries quantum oracle for $n(n+1)$ times and classical oracle for $2$ times. Thus the complexity of Algorithm 3 is $O(n^2)$. The distinguishing algorithm used in \cite{KL16,KM10,SS17} is based on Simon's algorithm and its complexity is only $O(n)$, which is better then ours. But our algorithm provides a new and inspirational approach to attack block ciphers. And by using our attack strategy, one can find more than one distinguisher. For example, we can also define $F(b,x)=P_2(x\oplus P_1(s_b))\oplus(b,\cdots,b)$. Then $F$ has the linear structure $(1\|P_1(s_0)\oplus P_1(s_1))\in U_F^{(1,\cdots,1)}$, and we can use it to construct another quantum distinguisher by similar way.

\subsection{Application to the Even-Mansour construction}
The Even-Mansour construction is a simple scheme which builds a block cipher from a public permutation \cite{EM97}. Suppose $P:\{0,1\}^n\rightarrow\{0,1\}^n$ is a permutation, the encryption function is defined as
$$
E_{k_1,k_2}=P(x\oplus k_1)\oplus k_2,
$$
where $k_1,k_2$ are the keys. Even and Mansour have proved that this construction is secure in the random permutation model up to $2^{n/2}$ queries. However, Kuwakado and Morii proposed a quantum attack which could recover the key $k_1$ based on Simon's algorithm. Our attack strategy is similar with theirs, we use BV algorithm instead of the Simon's algorithm.

In order to recover the key $k_1$, we first define the following function:
\begin{align*}
F:\{0,1\}^n&\rightarrow\{0,1\}^n \\
x&\rightarrow E_{k_1,k_2}(x)\oplus P(x).
\end{align*}
Given the oracle access of $E_{k_1k_2}(\cdot)$, it is easy to construct the oracle $O_F$ which computes $F$ on superpositions. Since $F(x)\oplus F(x\oplus k_1)=0$ for all $x\in \mathbb{F}_2^n$, $k_1$ is a linear structure of $F$, or more accurately, $k_1\in U_F^{(00\cdots 0)}$. Therefore, by running Algorithm 2 on $F$ with minor modification we can obtain $k_1$. Specifically, following algorithm can recover $k_1$ with a overwhelming probability.
\begin{framed}
\noindent
\textbf{Algorithm 4}

\noindent
The oracle access of $F=(F_1,\cdots,F_n)$ is given. Let $p(n)$ be an arbitrary polynomial function of $n$, and initialize the set $H:=\Phi$.\\
1 $\,\,$\textbf{For} $j=1,2,\cdots,n$, \textbf{do}\\
2 $\,$ \quad \textbf{For} $p=1,\cdots,p(n)$, \textbf{do}

\noindent
\hangafter 1
\hangindent 2.7em
3 $\,$\quad\quad Run the BV algorithm on $F_j$ to get a $n$-bit output $\omega\in N_{F_j}$.

\noindent
4 $\,\,$\quad\quad Let $H=H\cup\{\omega\}$.\\
5 $\,$\quad\textbf{End}\\
6 $\,$\quad Solve the system of linear equations $\{x\cdot \omega=0|\omega\in H\}$ to get $A_j^0$.\\
7 $\,$\quad  \textbf{If} $A_j^0\subseteq\{(0,\cdots,0)\}$,\textbf{then} output ``No'' and halt.\\
8 $\,$\quad\textbf{Else}, Let $H=\Phi$.\\
9 $\,\,$\textbf{End}.\\
10  \textbf{If} $A_1^0\cap\cdots\cap A_n^0\subseteq\{(0,\cdots,0)\}$, \textbf{then} output ``No'' and halt.

\noindent
\hangafter 1
\hangindent 1.4em
11 \textbf{Else} choose an arbitrary nonzero $a\in A_1^0\cap\cdots\cap A_n^0$ and output $a$.
\end{framed}
\begin{Theorem}
 Running Algorithm 4 with $n^2$ ($p(n)=n$) queries on $F$ gives the key $k_1$ except a negligible probability.
\end{Theorem}

\noindent
\textbf{\emph{Proof}}. By the similar proof of Lemma 3, we can obtain that for $j=1,\cdots,n$
\begin{equation}
\delta_{F_j}(k_1)\triangleq\frac{1}{2^n}\max_{\substack{\alpha\in \mathbb{F}_2^{n}\\ \alpha\notin\{(o,\cdots,0),k_1\}}}|\{x\in \mathbb{F}_2^n|F_j(x)=F_j(x\oplus \alpha)\}|\leq\frac{2}{3}
\end{equation}
holds except a negligible probability. Here, $\delta_{F_j}(k_1)$ is still a random variable since $F_j$ is determined by random functions $P_1,P_2$. Equation $(8)$ indicates that
$$
\delta'_F=\max_j\delta_{F_j}'\leq\frac{2}{3}
$$
holds except a negligible probability. Then according to Theorem 3, the probability that Algorithm 4 outputs $k_1$ is greater than $1-(\frac{2}{3})^n$, which completes the proof.

$\hfill{} \Box$

About the complexity of Algorithm 4, according to the equation $(8)$ and Theorem 2, the probability of $A_j^0$ containing the vectors apart from $k_1$ and $0$ is negligible. Thus finding the intersection of $A_j^0\,'s$ almost needs no calculation. In addition, Algorithm 4 needs to query quantum oracle for $n^2$ times. Thus its complexity is $O(n^2)$.

\section{Differential cryptanalysis}
In this section, we look at the linear structures from another view: the differentials of a encryption function. Based on this, we give three ways to execute differential cryptanalysis, which we call quantum differential cryptanalysis, quantum small probability differential cryptanalysis and quantum impossible differential cryptanalysis respectively. Unlike the classical differential cryptanalysis, the success probability of the first two methods is related to the key used for encryption algorithm. Specifically, suppose $q(n)$ is an arbitrary polynomial. For the first two methods, we can execute the corresponding attack algorithms properly so that they work for at least $(1-\frac{1}{q(n)})$ of the keys in the key space. While the third method works for all keys in the key space.

\subsection{Quantum differential cryptanalysis}
Differential cryptanalysis is a chosen-plaintext attack. Suppose $E:\{0,1\}^n\rightarrow\{0,1\}^n$ is the encryption function of a r-round block cipher. Let $F_k$ be the function which maps the plaintext $x$ to the input $y$ of the last round, where $k$ denotes the key of the first $r-1$ rounds. Let $F_k(x)=y$, $F_k(x')=y'$, then $\Delta x=x\oplus x'$ and $\Delta y=y\oplus y'$ are called the input difference and output difference respectively. The pair $(\Delta x,\Delta y)$ is called a differential. Differential cryptanalysis is composed by two phases. In the first phase, the attacker tries to find a high probability differential of $F_k$. In the second phase, according to the high probability differential that has been found, the attacker tests all possible candidate subkeys and then recover the key of the last round. Our algorithm is applied in the first phase, while a quantum algorithm is applied in the second phase in \cite{ZL15}.

Intuitively, we can use Algorithm 2 to find the high probability differentials of $F_k$. However, there exists a problem that the oracle access of $F_k$ is not available. The attacker can only query the whole encryption function $E$. In classical differential cryptanalysis, the attacker analyzes the properties of the encryption algorithm and searches for the high probability differentials that is independent of the key, i.e. the differentials that always have high probability no matter what the key is. We try to apply the same idea to our attack. But unfortunately, we still haven't found a way to obtain the key-independent high probability differentials of $F_k$ using BV algorithm. However, we can modify Algorithm 2 to find the differentials that have high probability for the most of keys. To do this, we treat the key as a part of the input of the encryption function and run Algorithm 2 on this new function. Specifically, suppose $m$ be the length of the key in the first $r-1$ rounds and $\mathcal{K}=\{0,1\}^m$ be corresponding key space. Define the following function
\begin{align*}
G:\{0,1\}^n\times\{0,1\}^m&\rightarrow\{0,1\}^n\\
   (\,x\,\,\,\,\,\,,\,\,\,\,\,\,\,k)\quad&\rightarrow F_k(x).
   \end{align*}
$G$ is deterministic and known to the attacker. Thus the oracle access of $G$ is available. (Actually, the oracle access of each $G_j$ is available.) By executing Algorithm 2 on $G$, one is expected to obtain a high probability differential of $G$ with overwhelming probability. But in order to make it also the differential of $F_k$, the last $m$ bits of the input difference, which corresponds to the difference of the key, needs to be zero. To do this, we modify the Algorithm 2 slightly as below:
 \begin{framed}
\noindent
\textbf{Algorithm 5}

\noindent
The oracle access of $G=(G_1,\cdots,G_n)$ is given. Let $p(n)$ be an arbitrary polynomial function of $n$, and initialize the set $H:=\Phi$.\\
1 \,\,\textbf{For} $j=1,2,\cdots,n$, \textbf{do}\\
2 \,\,\quad\textbf{For} $p=1,\cdots,p(n)$, \textbf{do}

\noindent
\hangafter 1
\hangindent 3.45em
3 \,\,\quad\quad Run the BV algorithm on $G_j$ to get a $(n+m)$-bit output $\omega=(\omega_1,\cdots,\omega_n,\omega_{n+1},\cdots,\omega_{n+m})\in N_{G_j}$.

\noindent
4 \,\,\,\,\quad\quad Let $H=H\cup\{(\omega_1,\cdots,\omega_n)\}$.\\
5 \,\,\quad\textbf{End}

\noindent
\hangafter 1
\hangindent 2.3em
6 \,\,\quad Solve the system of linear equations $\{x\cdot\omega =i_j|\omega\in H\}$ to get the set $A_j^{i_j}$ for $i_j=0,1$, respectively.

\noindent
7 \,\,\quad  \textbf{If} $A_j\triangleq A_j^0\cup A_j^1\subseteq\{(0,\cdots,0)\}$, \textbf{then} output ``No'' and halt.\\
8 \,\,\quad\textbf{Else}, Let $H=\Phi$.\\
9 \,\,\,\textbf{End}.

\noindent
10  \textbf{If} $A_1\cap\cdots\cap A_n\subseteq\{(0,\cdots,0)\}$, \textbf{then} output ``No'' and halt.

\noindent
\hangafter 1
\hangindent 1.4em
11 \textbf{Else} choose an arbitrary nonzero vector $a\in A_1\cap\cdots\cap A_n$ and output $(a,i_1,\cdots,i_n)$, where $i_1,\cdots,i_n$ is the superscript such that $a\in A_1^{i_1}\cap A_2^{i_2}$ $\cap\cdots\cap A_n^{i_n}$.
\end{framed}

By running Algorithm 5, one can find a differential of $F_k$ that has high probability for the most of keys. Specifically, we have following theorem:
\begin{Theorem}
   Suppose $q(n)$ is an arbitrary polynomial of $n$. If running Algorithm 5 with $np(n)$ quantum queries on $G$ gives a vector $(a,i_1,\cdots,i_n)$, then there exist a subset $\mathcal{K}'\subseteq\mathcal{K}$ such that $|\mathcal{K}'|/|\mathcal{K}|\geq1-\frac{1}{q(n)}$ and for all $k\in\mathcal{K}'$, it holds that
   \begin{equation*}
   Pr\Big[\frac{|\{x\in \mathbb{F}_2^n|F_k(x\oplus a)\oplus F_k(x)=i_1,\cdots,i_n\}|}{2^n}>1-\epsilon\Big]>\big(1-\exp(-\frac{2p(n)\epsilon^2}{q(n)^2n^2})\big)^n.
   \end{equation*}
\end{Theorem}

\quad\\
\vskip -0.17cm

\noindent
\textbf{\emph{Proof}}. Since $a\cdot(\omega_1,\cdots,\omega_n)=0$ indicates $(a\|0,\cdots,0)\cdot(\omega_1,\cdots,\omega_{n+m})=0$, the vector $(a\|0\cdots,0)$ can be seen as an output when we execute Algorithm 2 on $G$. According to Theorem 3,
\begin{equation}
\frac{|\{z\in \mathbb{F}_2^{n+m}|G(z\oplus(a\|0,\cdots,0)\oplus G(z)=i_1,\cdots,i_n\}|}{2^{n+m}}>1-n\epsilon_0
\end{equation}
holds with a probability greater than $(1-\exp(-2p(n)\epsilon_0^2))^n$. Let
$$
V(k)=\frac{|\{x\in \mathbb{F}_2^{n}|F_k(x\oplus a)\oplus F_k(x)=i_1,\cdots,i_n\}|}{2^{n}}.
$$
Equation $(9)$ indicates $\mathbb{E}_k(V(k))>1-n\epsilon_0$, where $\mathbb{E}_k(\cdot)$ means the expectation when the key $k$ is chosen uniformly at random from $\mathcal{K}$. Therefore, if the equation $(9)$ holds, for any polynomial $q(n)$, we have
$$
Pr_k[V(k)>1-q(n)n\epsilon_0]>1-\frac{1}{q(n)}.
$$
That is, for at least $(1-\frac{1}{q(n)})$ of keys in $\mathcal{K}$, it holds that $V(k)>1-q(n)n\epsilon_0$. Let $\mathcal{K}'$ be the set of these keys, then $|\mathcal{K}'|/|\mathcal{K}|\geq1-\frac{1}{q(n)}$, and for all $k\in\mathcal{K}'$, it holds that
\begin{equation*}
Pr\Big[V(k)>1-q(n)n\epsilon_0\Big]>\big(1-\exp(-2p(n)\epsilon_0^2)\big)^n.
\end{equation*}
The conclusion is obtained by letting $\epsilon=q(n)n\epsilon_0$.

$\hfill{} \Box$

According to Theorem 6, if $p(n)=O(n^3q(n)^2)$, then for any $k\in\mathcal{K}_1$ and any constant $c$, $V(k)>1-\frac{1}{c}$ holds except a negligible probability. For any constant $c_1,c_2$, if $p(n)=\frac{1}{2}c_1^2n^2q(n)^2\ln{(c_2n)}$, then for any $k\in\mathcal{K}_1$, we have
\begin{small}
\begin{equation*}
Pr\Big[\frac{|\{x\in \mathbb{F}_2^{n}|F_k(x\oplus a)\oplus F_k(x)=i_1,\cdots,i_n\}|}{2^{n}}>1-\frac{1}{c_1}\Big]>1-\frac{1}{c_2}.
\end{equation*}
\end{small}\\
\vskip -0.65cm

\noindent
When the attacker performs differential cryptanalysis, he or she first chooses constants $c_1,c_2$, then executes Algorithm 5 with $p(n)=\frac{1}{2}c_1^2n^2q(n)^2\ln{(c_2n)}$ to obtain a differential of $F_k$. The obtained differential has high probability for at least $(1-\frac{1}{q(n)})$ of keys in $\mathcal{K}$. Afterwards, the attacker determines the subkey in the last round according to this high probability differential, which can be done as in classical differential cryptanalysis. To analyze the complexity of Algorithm 5, we divide it into two parts: running BV algorithm to obtain the sets $A_j\,'s$; and finding the intersection of $A_j\,'s$. In the first part, Algorithm 5 needs to run BV algorithm for $np(n)=O(n^2q(n)^2\ln{n})$ times. Thus $O(n^2q(n)^2\ln{n})$ quantum queries are needed. As for the second part, the corresponding complexity depends on the size of the sets $A_j\,'s$. Suppose $t=\max_j|A_j|$, then the complexity of finding the intersection by sort method is $O(nt\log{t})$. The value of $t$ depends on the property of the encryption algorithm. Generally speaking, $t$ will not be large since a well constructed encryption algorithm usually does not have many linear structures. In addition, one can also choose a greater $p(n)$ to decrease the value of $t$. Therefore, the complexity of Algorithm 5 is $O(n^2q(n)^2\ln{n})$.

One of the advantages of our algorithm is that it can find the high probability differential directly. While in classical case, the attacker needs to analyze the partial structures of the encryption algorithm respectively and then seek the high probability differential characteristics, which may be much more complicated with the increase of the number of rounds.

\subsection{Quantum small probability differential cryptanalysis}
In this subsection we present a new way to execute differential cryptanalysis, which is called quantum small probability differential cryptanalysis. As shown in the previous sections, the way we find differentials of a vector function is to first search for the differentials of each component functions respectively, and then choose a public input difference and output the corresponding differential. Although this method will slightly increase the complexity of the attack algorithm, it may bring advantages in some applications. Quantum small probability differential cryptanalysis is such an example. Differential cryptanalysis using small-probability differentials was considered in \cite{BK97,KR99}. In differential analysis, the attacker needs to use the differentials with notable statistical properties to distinguish the block cipher from a random permutation, such as differential cryptanalysis and impossible differential cryptanalysis. As for the small probability differentials, since any differential of a random permutation only has a very small probability, the ``small-probability'' property of the entire differential cannot be directly used to distinguish the block cipher from a random permutation. However, if we consider each component function of the encryption function respectively, it will be possible to execute cryptanalysis based on small-probability differentials. Specifically, let $F_k:\{0,1\}^n\rightarrow\{0,1\}^n$ be the function which maps the plaintext $x$ to the input $y$ of the last round of the encryption algorithm, and $(\Delta x,\Delta y)$ is a differential of $F_k$ with small probability. For a random permutation $P=(P_1,\cdots,P_n)$, the differential $(\Delta x,\Delta y)$ appears with probability about $\frac{1}{2^n}$. But for the component function $P_j$, the probability of the differential $(\Delta x,\Delta y_j)$ appearing is $\frac{1}{2}$, which is not small at all. Our attack strategy is based on this fact. The detailed procedure is as follows:

\textbf{I. Finding small probability differential:} Let $G(x,k)=F_k(x)$ as defined previously and $\mathcal{K}$ denote the key space of the first $r-1$ rounds. The oracle access of $G$ is available. The attacker first chooses two polynomials $q(n),l(n)$ of $n$, then run Algorithm 5 with $np(n)=n^4l(n)^2q(n)^2$ ( $p(n)=n^3l(n)^2q(n)^2$ ) queries on $G$ to get an output $(a,i_1,\cdots,i_n)$. Let $b=(\bar{i_1},\cdots,\bar{i_n})$, where $\bar{i_j}=i_j\oplus 1$. Then $(a,b)$ is a small probability differential of $F_k$ for at least $(1-\frac{1}{q(n)})$ of keys in $\mathcal{K}$.

\textbf{II. Key recovering:} Suppose $\mathcal{S}$ is the set of all possible subkey of the last round. For each $s\in\mathcal{S}$, we set the corresponding counter $C_s$ to be zero and do as follows: fix the input difference $a$, and make $2l(n)^2$ classical queries on whole encryption function to get $2l(n)^2$ ciphers. Then decrypt the last round to obtain $l(n)^2$ output differences $\Delta y^{(1)},\cdots,\Delta y^{(l(n)^2)}$ of $F_k$. Let $\Delta y^{(i)}=(\Delta y_{1}^{(i)},\Delta y_{2}^{(i)},\cdots,\Delta y_{n}^{(i)})$. For $i=1,\cdots,l(n)^2$ and $j=1,\cdots,n$, if $\Delta y_{j}^{(i)}=b_j$, let the counter $C_s=C_s+1$. Afterwards, calculate the ratio $\lambda_s=C_s/nl(n)^2$. The attacker chooses the key $s\in\mathcal{S}$ which has the smallest ratio $\lambda_s$ to be the subkey of the last round.

To justify that above attack procedure work for at least $(1-\frac{1}{q(n)})$ of keys in $\mathcal{K}$, we give following theorem:
\begin{Theorem} There exists a subset $\mathcal{K}_1\subseteq\mathcal{K}$ such that:\\
\vskip -0.03cm

\noindent
\emph{(1)} $|\mathcal{K}_1|/|\mathcal{K}|\geq1-\frac{1}{q(n)}$;\\
\vskip -0.03cm

\noindent
\emph{(2)} If the key used for the first $r-1$ rounds of the encryption algorithm is in $\mathcal{K}_1$ and $s$ is the right subkey of the last round, then the ratio $\lambda_s$ obtained by the above procedure satisfies
$$
Pr[\,\, \lambda_s\geq \frac{1}{l(n)}\,\, ]\leq3\exp(-n/2).
$$
\end{Theorem}

\noindent
\textbf{\emph{Proof}}. According to Theorem 1 and the definition of $G$, for any $j=1,\cdots,n$,
\begin{equation}
\frac{|\{z\in \mathbb{F}_2^{m+n}|G_j(z)+G_j(z\oplus(a\|0))=b_j\}|}{2^{m+n}}\leq\epsilon
\end{equation}
holds with a probability greater than $1-\exp(-2p(n)\epsilon^2)$. Similar to the proof of Theorem 6, we let
$$
V_j(k)=\frac{|\{x\in \mathbb{F}_2^n|F_{kj}(x\oplus a)+F_{kj}(x)=b_j\}|}{2^n}.
$$
The equation $(10)$ indicates $\mathbb{E}_k(V_j(k))\leq\epsilon$, where $\mathbb{E}_k(\cdot)$ means the expectation when the key $k$ is chosen uniformly at random from $\mathcal{K}$. Therefore, if the equation $(10)$ holds, we have $Pr_k[V_j(k)\leq nq(n)\epsilon]\geq1-\frac{1}{nq(n)}$. In other words, for each $j\in\{1,\cdots,n\}$, $V_j(k)\leq nq(n)\epsilon$ holds for at least $(1-\frac{1}{nq(n)})$ of keys in $\mathcal{K}$. Then by similar analysis in the proof of Theorem 3, for at least $(1-\frac{1}{q(n)})$ of keys in $\mathcal{K}$, it holds that
$$
V_j(k)\leq nq(n)\epsilon, \text{\quad\quad} \forall j\in\{1,\cdots,n\}.
$$
Let $\mathcal{K}'$ be the set of these keys, then $|\mathcal{K'}|/|\mathcal{K}|\geq1-\frac{1}{q(n)}$, and for all $k\in\mathcal{K}'$, it holds that
\begin{equation*}
Pr\Big[\,\,\frac{|\{x\in \mathbb{F}_2^n|F_{kj}(x\oplus a)+F_{kj}(x)=b_j\}|}{2^n}\leq nq(n)\epsilon\,\,\Big]>1-\exp(-2p(n)\epsilon^2)
\end{equation*}
Let $\epsilon=\frac{1}{2nl(n)q(n)}$. Noticing $p(n)=n^3l(n)^2q(n)^2$, we have
\begin{equation}
Pr\Big[\,\,\frac{|\{x\in \mathbb{F}_2^n|F_{kj}(x\oplus a)+F_{kj}(x)=b_j\}|}{2^n}\leq \frac{1}{2l(n)}\,\,\Big]>1-\exp(-n/2).
\end{equation}
That is, for all $j=1,\cdots,n$,
\begin{equation}
Pr_x\Big[\,\,F_{kj}(x)+F_{kj}(x\oplus a)=b_j\,\,\Big]\leq\frac{1}{2l(n)}
\end{equation}
holds except a negligible probability. For $i=1,2,\cdots,l(n)^2$, $j=1,\cdots,n$, we define the random variable
$$
Y(i,j)=\left\{\begin{array}{cc}
1 & \quad\,\Delta y_{j}^{(i)}=b_j;\\
&\\
0 & \quad\,\Delta y_{j}^{(i)}\neq b_j.\\
\end{array}
\right.
$$
For every $i,j$, the equation $(12)$ indicates $\mathbb{E}_x(Y(i,j))\leq\frac{1}{2l(n)}$ except a negligible probability. (Here $\mathbb{E}_x$ means the expectation when output difference is obtained by choosing plaintext $x$ uniformly at random. $\mathbb{E}_x(Y(i,j))$ is still a random variable since it is a function of the vector $a$, which is a random variable output by Algorithm 5.) According to Hoeffding's inequality and the fact that the equation $(12)$ holds except a probability $\exp(-n/2)$, it holds that
$$
Pr\Big[\frac{\sum_{i,j}Y(i,j)}{nl(n)^2}\geq\frac{1}{2l(n)}+\delta\Big]\leq 2\exp(-2nl(n)^2\delta^2)+\exp(-n/2).
$$
Let $\delta=\frac{1}{2l(n)}$. Noticing $\sum_{i,j}Y(i,j)/nl(n)^2=\lambda_s$, we have
$$
Pr[\lambda_s\geq\frac{1}{l(n)}]\leq3\exp(-n/2),
$$
which completes the proof.

$\hfill{} \Box$

In key recovering phase, the attacker computes $l(n)^2$ output difference to get the ratio $\lambda_s$ for every $s\in\mathcal{S}$. If $s$ is not the right key of the last round, the $l(n)^2$ differentials $(a,\Delta y^{(i)})$ can be seen as differentials of a random permutation. Then the probability of $Y(i,j)=1$ is approximate to $\frac{1}{2}$ for every $i,j$. Therefore, the expectation of $\lambda_s$ is approximate to $\frac{1}{2}$. On the other hand, if $s$ is the right key, the probability of $\lambda_s\geq\frac{1}{l(n)}$ is negligible according to Theorem 7. This notable difference makes our attack strategy feasible for at least $(1-\frac{1}{q(n)})$ of keys in $\mathcal{K}$. About the complexity of the attack procedure, there are $n^4l(n)^2q(n)^2$ quantum queries and $2l(n)^2$ classical queries needed in total.

The basic idea of quantum small probability differential cryptanalysis is similar to the idea of quantum differential cryptanalysis, that is, using some notable statistical difference to distinguish a encryption function from a random permutation. The main difference of these two methods is in key recovering phase. In quantum differential cryptanalysis, the attacker treats the differential as a whole and records the number of times it appears, while in quantum small probability differential cryptanalysis, the attacker considers every bit of the output differences respectively and records the number of times they appear.

\subsection{Quantum impossible differential cryptanalysis}

Impossible differential cryptanalysis is also a chosen-plaintext attack. Suppose $F_k:\{0,1\}^n\rightarrow\{0,1\}^n$ and the key space $\mathcal{K}$ are defined as before. A differential $(\Delta x,\Delta y)$ is called a impossible differential of $F_k$ if it satisfies that
$$
F_k(x\oplus\Delta x)+F_k(x)\neq \Delta y, \quad\forall x\in \mathbb{F}_2^{n}.
$$
Impossible differential cryptanalysis is composed by two phases. In the first phase, the attacker tries to find an impossible differential $(\Delta x,\Delta y)$ of $F_k$. And in the second phase, the attacker uses the found impossible differential to sieve the subkey of the last round. Specifically, the attacker fixes the input difference $\Delta x$, and make classical queries on the whole encryption function to get a certain number of ciphers. Then for any possible key $s$ of the last round, the attacker uses it to decrypt these ciphers and obtains corresponding output differences of $F_k$. If $\Delta y$ appears among these output differences, then the attacker rules out $s$. Our algorithm is applied in the first phase.

Let $G(x,k)=F_k(x)$ as defined previously. The oracle access of $G$ is available. A algorithm to find the impossible differentials of $F_k$ is as follows:
 \begin{framed}
\noindent
\textbf{Algorithm 6}

\noindent
The oracle access of $G=(G_1,\cdots,G_n)$ is given. Let $p(n)$ be an arbitrary polynomial function of $n$, and initialize the set $H:=\Phi$.\\
1 $\,\,$\textbf{For} $j=1,2,\cdots,n$, \textbf{do}\\
2  \quad \textbf{For} $p=1,\cdots,p(n)$, \textbf{do}

\noindent
\hangafter 1
\hangindent 3em
3 \quad\quad Run the BV algorithm on $G_j$ to get a $(n+m)$-bit output $\omega=(\omega_1,\cdots,\omega_n,\omega_{n+1},\cdots,\omega_{n+m})\in N_{G_j}$.

\noindent
4 \quad\quad Let $H=H\cup\{(\omega_1,\cdots,\omega_n)\}$.\\
5 \quad\textbf{End}

\noindent
\hangafter 1
\hangindent 2em
6 \quad Solve the system of linear equations $\{x\cdot \omega=0|\omega\in H\}$ to get the set $B^1$ and $\{x\cdot \omega=1|\omega\in H\}$ to get the set $B^0$. Let $B_j= B_j^0\cup B_j^1$.

\noindent
\hangafter 1
\hangindent 2em
7 \quad  \textbf{If} $B_j\supsetneq\{(0,\cdots,0)\}$, \textbf{then} choose an arbitrary nonzero $a\in B_j$ and output $(j,a,i_j)$, where $i_j$ is the superscript such that $a\in B_j^{i_j}$, and halt.

\noindent
8 \quad\textbf{Else}, Let $H=\Phi$.\\
9 $\,\,$\textbf{End}.\\
10 Output ``No''.
\end{framed}

Suppose the attacker gets $(j,a,i_j)$ by running Algorithm 6 with $p(n)=O(n)$. Let $\delta'_G=\max_{1\leq j\leq n}\delta'_{G_j}$. If $\delta'_G\leq p_0<1$ for some constant $p_0$, then according to Theorem 2, $(a,\times,\cdots,\times,i_j,\times,\cdots,\times)$ will be an impossible differential of $F_k$ except a negligible probability. Here $``\times''$ means the corresponding bit can be either 0 or 1. Specifically, we have following theorem:
\begin{Theorem}
If $\delta'_G\leq p_0<1$ and running Algorithm 6 with $np(n)=n^2$ ($p(n)=n$) queries on $G$ gives a vector $(j,a,i_j)$, then for any key $k\in\mathcal{K}$ and any $i_1,\cdots,i_{j-1}$, $i_{j+1},\cdots,i_n\in\{0,1\}$, it holds that
\begin{equation}
F_k(x)\oplus F_k(x\oplus a)\neq(i_1,\cdots,i_{j-1},i_j,i_{j+1},\cdots,i_n),\quad\forall x\in \mathbb{F}_2^n
\end{equation}
except a negligible probability. That is, $(a,i_1,\cdots,i_n)$ is a impossible differential of $F_k$ except a negligible probability.
\end{Theorem}

\noindent
\textbf{\emph{Proof}}. According to Theorem 2, we have $Pr[a\in U_{G_j}^{\bar{i_j}}]>1-p_0^n$. Thus $Pr[G_j(z)\oplus G_j(z\oplus(a\|0))\neq i_j,\forall z\in \mathbb{F}_2^{m+n}]>1-p_0^n$. This indicates for all $k\in\mathcal{K}$,
$$
Pr[F_{kj}(x)\oplus F_{kj}(x\oplus a)\neq i_j,\forall x\in \mathbb{F}_2^n]>1-p_0^n.
$$
Since the probability that the equation $(13)$ holds is no less than the above probability, the conclusion holds.

$\hfill{} \Box$

From Theorem 8, we can see that by running Algorithm 6 with $O(n^2)$ queries the attacker may find impossible differentials of $F_k$. Unlike the other two kinds of differential cryptanalysis proposed in previous two subsections, the ``impossibility'' of the found differential holds for all keys in $\mathcal{K}$. But Algorithm 6 can only find the impossible differentials whose ``impossibility'' concentrates on a certain bit. In other words, only when there exists some $j$ such that $F_{kj}$ has impossible differentials, can Algorithm 6 find impossible differentials of $F_k$. Although our algorithm can only find such special impossible differentials, it still provide a new and inspirational approach for impossible differential cryptanalysis. In addition, one of the main shortcomings of traditional impossible differential cryptanalysis is the difficulties in extending the differential path, which limits the number of rounds that can be attacked. Our approach does not have this problem since it treats the first $r-1$ rounds as a whole.

\section{Discussion and conclusion}

In this paper, we construct new quantum distinguishers for the 3-round Feistel scheme and propose a new quantum algorithm to recover partial key of Even-Mansour construction. Afterwards, by observing that the linear structures of a encryption function are actually high probability differentials of it, we propose three ways to execute differential cryptanalysis. The quantum algorithms used for these three kinds of differential cryptanalysis all have polynomial running time. We believe our work provides some helpful and inspirational methods for quantum cryptanalysis.

There are many directions for future work. First, is it possible to modify the algorithms used for quantum differential analysis and quantum small probability differential cryptanalysis so that they can work for all keys in the key space. Also, under the premise of not affecting the success probability, how to reduce the complexity of our attacks is worthy of further study. In addition, all algorithms proposed in this article find differentials of a vector function by first searching for the differentials of its component functions respectively. There may exist other ways that find differentials of a vector function directly.

\begin{acknowledgements}
This work was supported by National Natural Science Foundation of China (Grant No.61672517), National Cryptography Development Fund (Grant No. MMJJ201 70108) and the Fundamental theory and cutting edge technology Research Program of Institute of Information Engineering, CAS (Grant No. Y7Z0301103).
\end{acknowledgements}



\begin{appendix}
\section{Proof of Theorem 1}

In this section we present the proof of Theorem 1, which can be found in \cite{LY16}. Theorem 1 is stated as following:

\noindent
\textbf{Theorem 1.}If running Algorithm 1 on a function $f\in\mathcal{B}_n$ gives sets $A^0$ and $A^1$, then for all $a\in A^i$ ($i=0,1$), all $\epsilon$ satisfying $0<\epsilon<1$, we have
\begin{equation*}
Pr(1-\frac{|\{x\in \mathbb{F}_2^n|f(x\oplus a)+f(x)=i\}|}{2^n}<\epsilon)>1-\exp(-2p(n)\epsilon^2).
\end{equation*}

\noindent
\noindent
\textbf{\emph{Proof}}. For all $a\in A^i$ ($i=0,1$),
$$
Pr_x[f(x\oplus a)+f(x)=i]=\frac{|\{x\in \mathbb{F}_2^n|f(x\oplus a)+f(x)=i\}|}{2^n}=\frac{|V_{f,a}^i|}{2^n}.
$$
Let $p=|V_{f,a}^i|/2^n$ and $q=1-p$, then $p,q\in[0,1]$. We define a random variable $Y$ as following:
\begin{equation*}
Y(\omega)=\left\{\begin{array}{cc}
0, & \quad\omega\cdot a=i;\\
1, & \quad\omega\cdot a\neq i.\\
\end{array}
\right.
\end{equation*}
According to Lemma 1, the expectation of $Y$ is $\mathbb{E}(Y)=1\cdot q=1-p$. The $p(n)$ times of running the BV algorithm produce $p(n)$ independent identical random variables $Y_1,\cdots,Y_{p(n)}$. By Hoeffding's inequality,
\begin{equation*}
Pr[\,\,q-\frac{1}{p(n)}\sum_{j=1}^{p(n)}Y_j\geq\epsilon\,\,]\leq \exp(-2p(n)\epsilon^2).
\end{equation*}
Note that $a\in A^i$, we have $\sum_jY_j$ must be 0 (otherwise there exists some $Y_j=1$, then $a\notin A^i$). Thus $Pr[q\geq\epsilon]\leq \exp(-2p(n)\epsilon^2)$. This indicates
$$
Pr[1-p<\epsilon]=Pr[q<\epsilon]>1-\exp(-2p(n)\epsilon^2),
$$
which completes the proof.

$\hfill{} \Box$

\end{appendix}
\end{document}